\documentclass[a4paper]{article}  

\pdfoutput=1

\usepackage{indentfirst}  
\usepackage{amssymb}  
\usepackage{bm}    
\usepackage{graphicx}  
\usepackage{url}
\usepackage{fancyhdr}

\usepackage{geometry}
\begin{document}
\thispagestyle{empty} \vspace*{0.8cm}
\hbox to\textwidth{\vbox{\hfill\noindent \\ \textit{Proceedings of the 8th
International Conference on Pedestrian and Evacuation Dynamics (PED2016)\\
Hefei, China - Oct 17 -- 21, 2016\\ Paper No. 18} \hfill}}
\par\noindent\rule[3mm]{\textwidth}{0.2pt}\hspace*{-\textwidth}\noindent
\rule[2.5mm]{\textwidth}{0.2pt}

\begin{center}
\textbf{\LARGE Influence of selfish and polite behaviours on a pedestrian
evacuation through a narrow exit: A quantitative characterisation}
\par\end{center}{\LARGE \par}

\begin{center}
{\LARGE Alexandre }{\LARGE NICOLAS}{\LARGE }%
\footnote{Now at: LPTMS (Bat 100), CNRS \& Universit\'e Paris-Sud, Orsay,
France.%
}{\LARGE , Sebasti\'{a}n }{\LARGE BOUZAT}{\LARGE{} and Marcelo N.
}{\LARGE KUPERMAN}
\par\end{center}{\LARGE \par}

\begin{center}
CONICET and Centro At\'{o}mico Bariloche, Bariloche, Argentina
\par\end{center}

{\LARGE \vspace*{2mm}
}{\LARGE \par}

\begin{center}
{\LARGE }%
\begin{minipage}[c]{15.5cm}%
\textbf{\emph{Abstract}}

We study the influence of selfish \emph{vs}. polite behaviours on
the dynamics of a pedestrian evacuation through a narrow exit. To
this end, experiments involving about 80 participants with distinct
prescribed behaviours are performed; reinjection of participants into
the setup allowed us to improve the statistics. Notwithstanding the
fluctuations in the instantaneous flow rate, we find that a stationary  regime
is almost immediately reached. The average flow rate increases monotonically
with the fraction $c_{s}$ of vying (selfish) pedestrians, which corresponds
to a ``faster-is-faster'' effect in our experimental conditions;
it is also positively correlated with the average density of pedestrians
in front of the door, up to nearly close-packing. At large $c_{s},$
the flow displays marked intermittency, with bursts of quasi-simultaneous
escapes.

In addition to these findings, we wonder whether the effect of cooperation
is specific to systems of intelligent beings, or whether it can be
reproduced by a purely mechanical surrogate. To this purpose, we consider
a bidimensional granular flow through an orifice in which some grains
are made \textquotedblleft{}cooperative\textquotedblright{} by repulsive
magnetic interactions which impede their mutual collisions.

\begin{minipage}[t]{13.1cm}%
\textbf{Keyworks} pedestrian, evacuation, behaviour, experimental,%
\end{minipage}%
\end{minipage}
\par\end{center}{\LARGE \par}

\section{Introduction}

In Boris Vian's \emph{Froth on the Daydream}, the main character witnesses
a massive collision on an ice skating rink; no sooner have the skaters
tumbled into a deadly pileup than housekeepers enter the rink and
sweep away their corpses while singing. In real life, crowd disasters
are not taken so light-heartedly; recent examples during the emergency
evacuation of a nightclub in Bucharest (October 2015) \cite{guardian2015bucharest}
or the massive stampede in Mina during the Muslim Hajj (September
2015) \cite{nytimes2015stampede}, which resulted in thousands of
casualties, demonstrate how tragic these events can be. The issue
of evacuation in public facilities and, more generally, questions
related to pedestrian flows are therefore of paramount importance.

It is needless to say, though, that the description or modelling of
the dynamics of crowds is, at the very least, challenging. Since first-principles
approaches are hopeless in any conceivable future, models have to
rely on a combination of assumptions and empirical observations. Fortunately,
the success of rather simple physical models in describing increasingly
complex entities (complex fluids, bacteria \cite{marchetti2013hydrodynamics},
birds and fish \cite{Gautrais2012deciphering}) bolsters the desire
for similarly successful models for pedestrian crowds. \emph{But how
detailed, or ``realistic'', need these models be to appropriately
describe the flow of pedestrians in a given situation?} The numerous
existing models and simulation software \cite{gwynne1999review,bellomo2011modeling}
vary greatly as regards their sophistication, from minimalistic cellular
automata \cite{Burstedde2001simulation} to continuous agent-based
models incorporating a large number of parameters. Even if we explicitly
bypass the psychological intricacy of the human mind and take for
granted the pedestrians' choice of a target point and his or her behaviour,
the dynamics that result from these choices, and how simple or complex
they are, remain mostly unknown.

The flow of pedestrians at a constriction, \emph{e.g.}, the disordered
egress through a narrow exit, seems to be largely governed by physical
constraints and is therefore worth studying from a mechanical perspective
(not to mention its importance for safety science). Still, one can expect
the attitude of pedestrians to play an important role in the dynamics.

Our objectives are two-fold. First, we wish to characterise the influence
of prescribed polite (cooperative) and selfish (vying) behaviours,
regardless of their psychological origins, on the evacuation process
of a heterogeneous crowd. Second, we investigate whether the observed
characteristics are tied to the inherent complexity of intelligent
beings or whether they can be reproduced in a simple granular system.
To this end, repulsive magnetic interactions are introduced between
some grains to mimic the avoidance of physical contact presumably
associated with polite behaviour.

We have chosen to divide the presentation of our work as follows.
In this contribution, we shall expose the underpinning of the granular
analogy and its ``behavioural extension'', present our original
setups in detail and expose the global flow properties of pedestrians
and grains. The \emph{microscopic} characterisation of the flows and a more
thorough
analysis of the granular analogy shall be published separately, in
a companion paper \cite{nicolas2016next}.

\section{Underpinning of the work}

In the last decade, large-scale experiments of pedestrian flows through
bottlenecks
have been conducted. Within the framework of a German collaborative
project, the flow of voluntary pedestrians under normal (cooperative)
conditions was studied at the abrupt narrowing of a corridor, from
$4\,\mathrm{m}$ to a variable width
$w\in\left[0.8\mathrm{m},1.2\mathrm{m}\right]$ \cite{seyfried2010enhanced}.
Contrary to perhaps less exhaustive
earlier measurements \cite{hoogendoorn2005pedestrian}, the flow was
observed to increase monotonically with $w$, instead of undergoing
a stepwise growth. For $w\gtrsim0.9\mathrm{m},$ a ``zipper effect''
is visible, whereby two lanes form in the narrower corridor and entrances
tend to be grouped in pairs, one pedestrian on either lane, with a
short headway to avoid physical contact; this, claim the authors,
leads to anticorrelated time headways $\Delta t$ between successive
passages. More recently, a Spanish-Argentine collaboration performed evacuation
drills through a narrow (69cm-wide)
door with almost 100 participants and revealed robust statistical
features. In each realisation, a given degree of competitiveness was
imposed on the whole crowd, from low (\emph{i.e.,}, avoidance of all
contacts) to high (mild pushes) \cite{Garcimartin2014experimental,Pastor2015experimental}.
These experiments provided evidence for the so called
``faster-is-slower'' effect, whereby a crowd made of individuals
eager to move faster actually needs longer to evacuate than a less
competitive crowd, due to the formation of clogs. Furthermore, the
distributions of time lapses $\Delta t$ between successive egresses
were found to be heavy-tailed and well described by power laws at
large $\Delta t$, viz.
\begin{equation}
p(\Delta t)\sim\Delta t^{-\alpha}.\label{eq:power_law_p}
\end{equation}
These features, along with a couple of others, appear to be universal
to flows of particles through narrow bottlenecks, as they have also
been reported for sheep entering a barn, granular hopper flows, and
colloids flowing through an orifice \cite{Zuriguel2014clogging,garcimartin2015flow}.
This analogy between pedestrian evacuations and granular flows through
a bottleneck suggests that the dynamics of pedestrians in such situations
are governed by the following few key mechanisms:

\emph{Excluded volume}. Clogging occurs in granular hopper flows because
of the formation of arches bridging the gap between walls at the outlet;
these arches resist pressure thanks to the solid contacts between
the grains that constitute them \cite{To2001jamming}. For orifices
smaller than 3 or 4 particle diameters in two dimensions (2D), permanent
clogs almost surely occur and the flow halts \cite{To2001jamming,janda2009unjamming}.
To et al. rationalised the probability of permanent clogging in 2D
by calculating the probability of formation of a convex arch between
the walls. In addition, contrary to the case of liquids, anisotropic
force chains develop in the granular system, so that the height of
the granular column is not transmitted down to the bottom; most strikingly,
the pressure measured at the outlet is virtually independent of the
height of the granular layer; this is the so called Janssen effect
\cite{janssen1895versuche}.

\emph{Fluctuations or vibrations. }Arch-like structures are also observed
in pedestrian flows through bottlenecks, but they lead to non-persistent
(although potentially long) clogs, responsible for the existence of
relatively large time lapses $\Delta t$ between successive egresses
and intermittency \cite{Pastor2015experimental}. In granular flows,
too, arches can be broken provided that the system is vibrated strongly
enough \cite{janda2009unjamming}. Therefore, in a model, such fluctuations
should be present in order to avoid permanent clogs; they are generally
described by a noise term.

As the orifice size $D_{0}$ is increased, at a given small vibration
amplitude, the importance of vibrations is reduced as the flow transits
from a regime of intermittent flow dominated by the time needed for
vibrations to break clogs, to a regime of continuous flow. The latter
regime is the usual range of applicability of Beverloo's relation
between the flow rate and $D_{0}$ \cite{beverloo1961flow,janda2009unjamming}.
A parameter controlling the transition between these regimes is $\Phi$,
the fraction of time during which the system flows \cite{Zuriguel2014clogging}.
Note that, according to Eq.~\ref{eq:power_law_p}, $\Phi\rightarrow0$
if $\alpha<2$. For pedestrian flows, as long as the door is wide
enough for agents to go through, $\Phi>0$.

\emph{Frictional contacts vs. no contacts.} In competitive evacuations,
people will be in contact. Conversely, under normal conditions collisions
are avoided and contacts are rare. One may reasonably think that this
lack of contacts will strongly affect the flow \cite{lumay2015flow}.
Indeed, in granular systems, simply replacing frictional grains with
frictionless ones reduces the probability of clogging, insofar as
arches become less stable \cite{To2001jamming}. For slighlty deformable
frictionless particles (surfactant-coated emulsion droplets), clogging
only occurs at very narrow outlets, only slightly larger than the
particle diameter \cite{hong2015jamming}. To go beyond frictionless
contacts and really investigate systems with no particle contacts,
Lumay et al. recently studied the 2D hopper flow of repulsive magnetic
discs \cite{lumay2015flow}. It is noteworthy that even without contacts
clogs were observed for aperture sizes significantly larger than the
particle diameter, through the formation of arches held by magnetic
repulsions. But, in the absence of a vibrating device in the setup,
the authors rather focused on wider apertures. Thus, the following
question is still open: \emph{does the preclusion of contacts between
grains in a hopper flow suffice to reproduce the change from a competitive
evacuation through a narrow door to a cooperative one?}

Overall, the aforementioned experiments deal with relatively homogeneous
crowds or systems. Here, in order to better illuminate the effect
of individual behaviours, we shall consider heterogeneous systems.

\section{Description of the setups}

\subsection{Pedestrian evacuation experiments}

\subsubsection*{Setup}

\begin{figure}
\begin{centering}
\includegraphics[width=4.5cm]{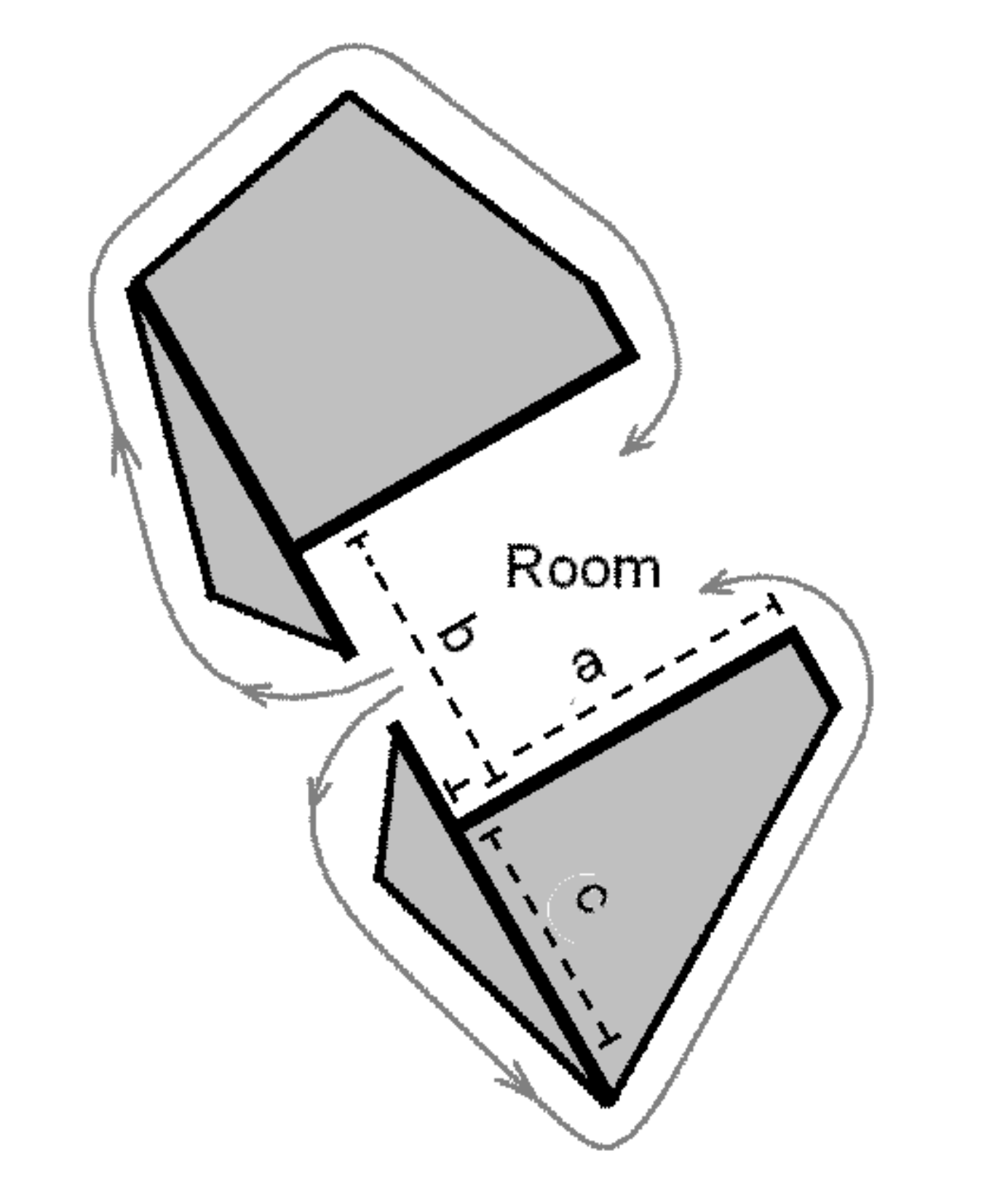}\includegraphics[width=7cm]{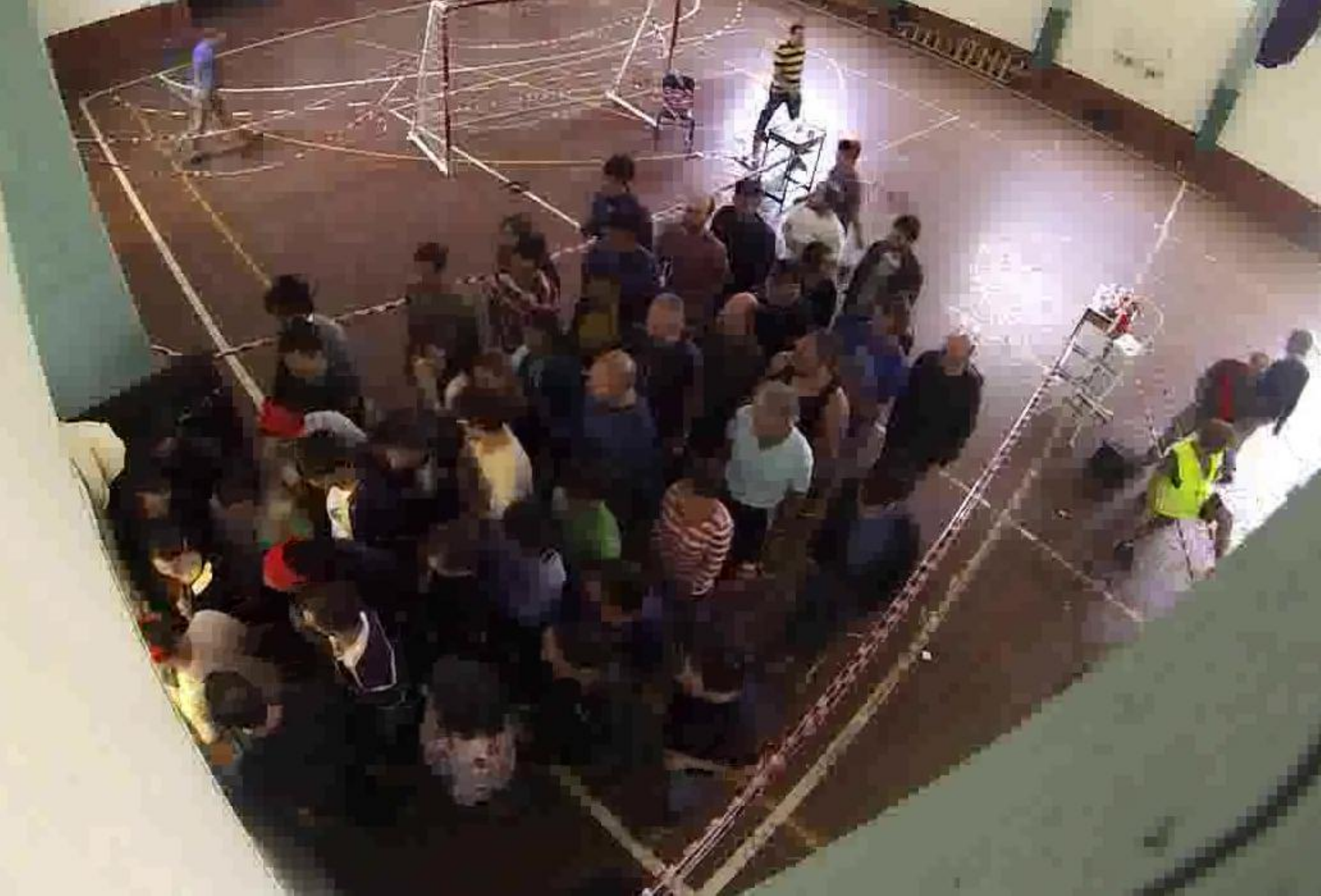}
\par\end{centering}

\caption{\label{fig:Geometry}(\emph{Left}) Sketch of the experimental
geometry, with $a=7.3\,\mathrm{m}$, $b=3.5\,\mathrm{m}$, and
$c=4.3\,\mathrm{m}$.
(\emph{Right}) snapshot of an evacuation.}

\end{figure}

We performed evacuation experiments involving more than 80 voluntary
participants on the premises of Centro At\'omico Bariloche (CAB), Argentina.
The participants were students and researchers, aged 20 to 55 for
the greatest part, with a woman/man ratio of about 1:3. They were
asked to evacuate a delimited space through a $72\,\mathrm{cm}$-wide door, with
the specific instructions
exposed below. The geometry of the setup is sketched in Fig.~\ref{fig:Geometry}.

In light of its potential risks, the experiment was validated beforehand
for ethical issues and prepared in collaboration with the Safety
and Hygiene group of CAB. The doors and the walls were
protected with paddings and the whole evacuation process was supervised
by 3 staff members who could stop it at any moment by blowing a whistle.
No incident whatsoever was reported during or after the drill.

This experimental setup is directly inspired from \cite{Garcimartin2014experimental,Pastor2015experimental},
but some significant changes were introduced. All participants were
instructed to ``head for the door purposefully (\emph{con ganas}),
but without running, pushing or hitting others'', but in each evacuation
drill, a fraction $c_{s}$ of participants was randomly selected to
behave selfishly while the rest should behave politely. Selfish participants
were told to wear a red headscarf to be recognisable on the videos
and allowed to ``elbow their way through the crowd, with mild contacts
but no violence whatsoever''. Their polite counterparts, on the other
hand, were to ``avoid any contact and try to keep their distance''.

Moreover, an innovative trick was devised to curb finite-size effects:
egressing participants were reinjected into the room, as if there were periodic
boundary conditions. To maintain homogeneity in the system and limit
the clustering of, say, fast participants, two reinjection circuits
were set up, a short one and a longer one, and the evacuees were directed
to either one alternatively.

These experiments were briefly described in (China)

\subsubsection*{Data analysis}

The flow at the exit was filmed by two video cameras (acquisition
rate: 60Hz) placed just above the door while a third camera provided
a more global view.

To study the dynamics at the individual scale, time frames of escapes
were constructed, following
Refs.~\cite{Garcimartin2014experimental,Pastor2015experimental},
by extracting a line of pixels just past the door from every frame
of the video and stitching these lines together: an example of such
a time frame is shown in Fig.~\ref{fig:Time-frames}. The individual
escape times were collected semi-manually by clicking on each pedestrian,
with the help of a home-made computer routine. The reproducibility of the
experimental results
was tested by performing two runs in nominally identical conditions, but at
different times and with different individuals as selfish agents;
the results were in fair agreement (7\% difference in the global flow
rate, similar - though not identical - distributions of time lapses
between escapes, \emph{data not shown}).

\subsection{Two-dimensional granular hopper flow}

\subsubsection*{Setup}

\begin{figure}
\begin{centering}
\includegraphics[width=7cm]{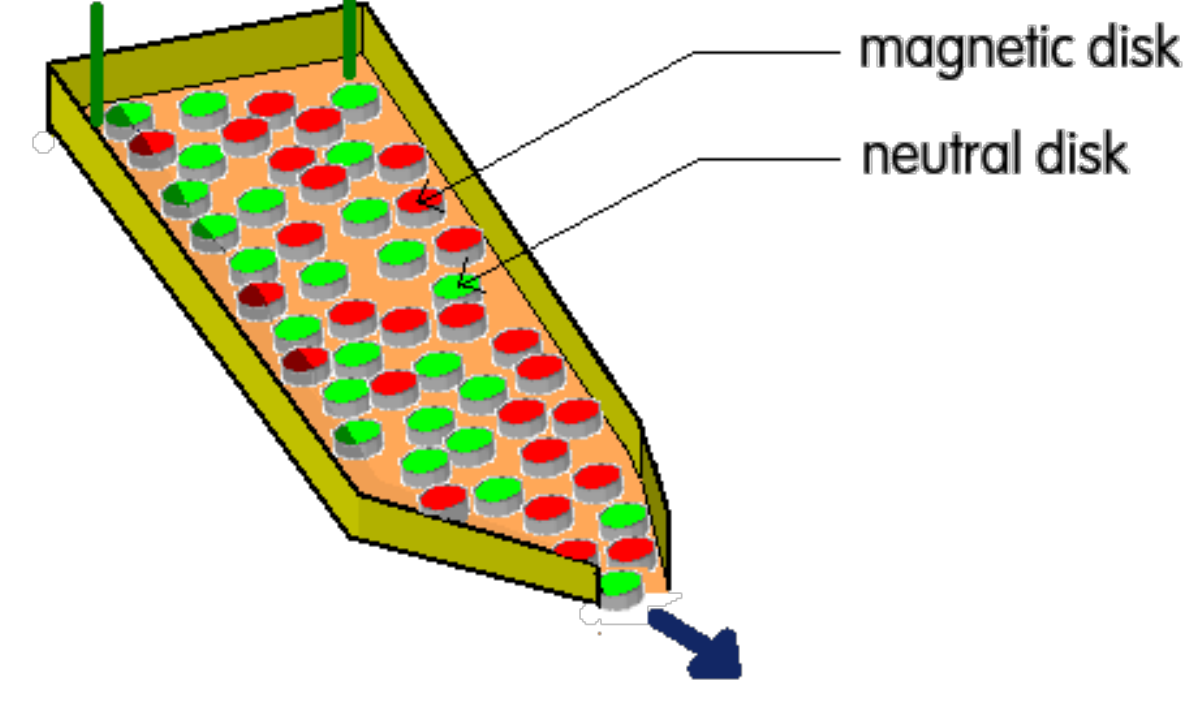}~
\includegraphics[width=5.5cm]{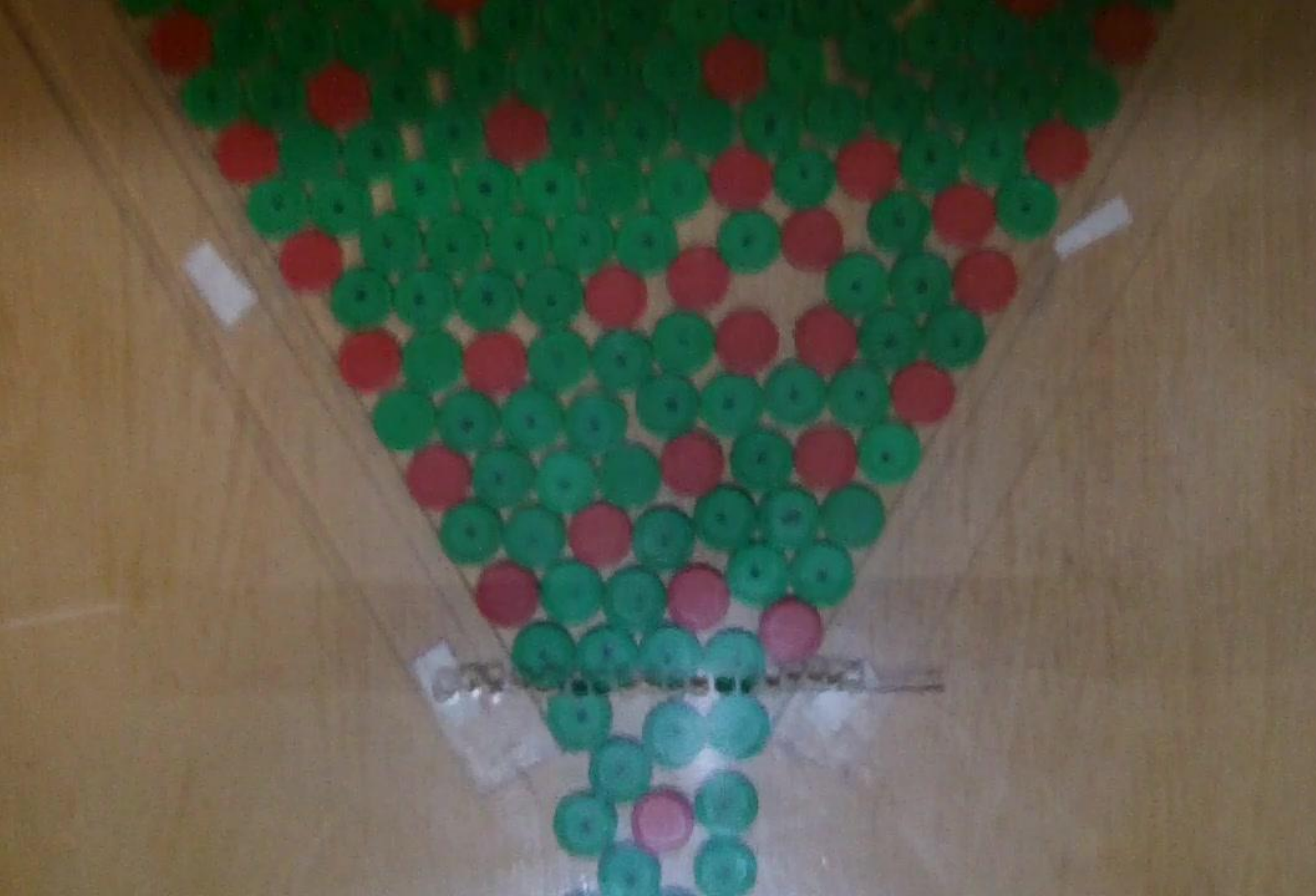}
\caption{\label{fig:Geometry-gran}(\emph{Left}) Sketch of the experimental
geometry and (\emph{right}) snapshot of an experiment. The chipwood
panel is inclined by about $18^{\circ}$ and vibrates.}

\par\end{centering}

\end{figure}

To investigate the analogy between pedestrian and granular flows,
we also conducted experiments on constricted flows of non-magnetic
(green) and magnetic (red) granular discs (see Fig.~\ref{fig:Geometry-gran}).
The discs are about $1\,\mathrm{mm}$-thick and confined between a vibrated
chipwood panel at the bottom and a plastic sheet. Thin plastic bars
placed between the plates delimit a 2D funnel-like region, with an
aperture of variable width $w=$3, 4 or 5cm and an opening angle of
$58^{\circ}$. 

Both types of discs consist of a circular plastic cap of 13mm of diameter,
mounted on a narrower bronze washer for the non-magnetic ones, or
a narrower neodyme magnet otherwise. Their total weights are
$0.545\mathrm{g}\pm0.01\mathrm{g}$
and $0.582\mathrm{g}\pm0.01\mathrm{g}$, respectively. The chipwood
panel was carefully polished and varnished to reduce the friction
coefficient and make it less heterogeneous. At rest, \emph{i.e.},
without vibrations, the static friction coefficient was about
$\mu\simeq0.32=\tan\left(18^{\circ}\right)$.
The chipwood panel was inclined by an angle roughly equal to the friction
angle, $18^{\circ}$, or perhaps slightly smaller, but the vibrations
produced by an unbalanced motor attached to the panel allowed sliding.
At the beginning of each experiment, the discs were inserted from
the top of the setup, in a relatively homogeneous fashion.

In the following, the aperture width $w$ and the relative fraction
of magnetic discs in the system are varied; all other parameters are
kept constant.

\subsubsection*{Data analysis}

\begin{figure}
\begin{centering}
\includegraphics[width=14cm]{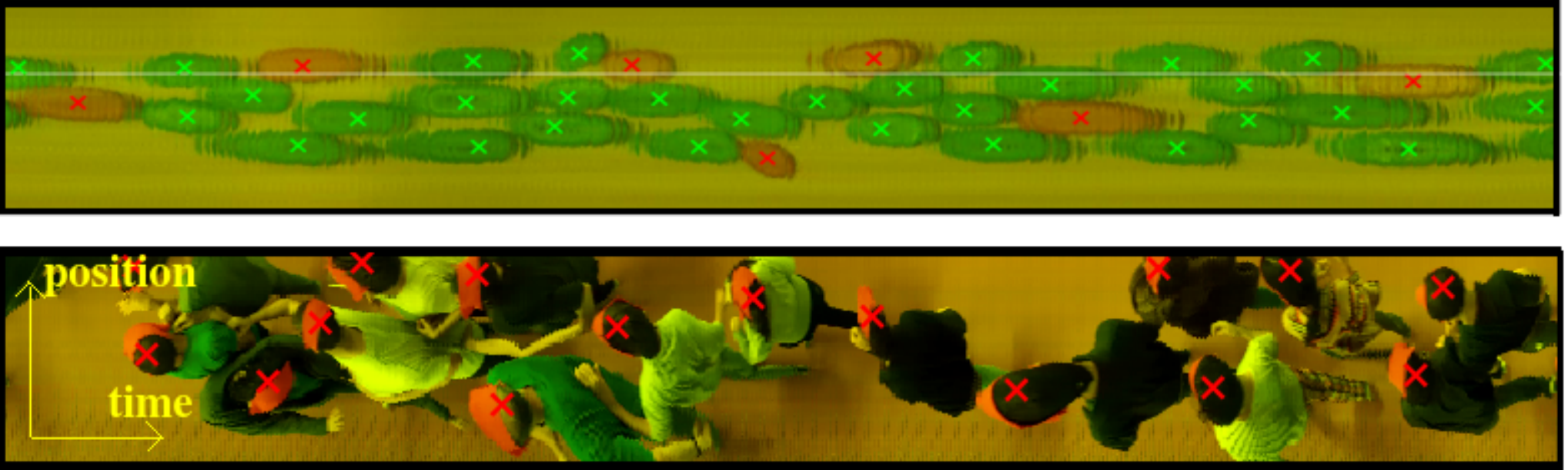}
\par\end{centering}

\caption{\label{fig:Time-frames}Time frames (zoom) for a granular experiment
(\emph{top}) and a pedestrian evacuation experiment (\emph{bottom}).
The vertical axis represents the position along the direction of the
door and the horizontal coordinate is time.}
\end{figure}

The granular flow was filmed with a 60Hz camera placed above the aperture.
The analysis is very similar to that performed for the pedestrian
experiments (see Fig.~\ref{fig:Time-frames}).

\section{Experimental results for the pedestrian evacuation}

Let us start with the study of the pedestrian flow. We recall that
the participants are prescribed either a polite or a selfish behaviour.
Here, we present results with $c_{s}=$0\%, 10\%, 60\%, 90\%, and
100\% of selfish participants. Generally, we only have data for one
run for each $c_{s}$, but, thanks to the recirculation, each run
comprises about 200 to 300 passages through the door.

\subsection{General observations}

As expected, no collisions or contacts are observed with crowds of
polite agents ($c_{s}=0\%$). On the contrary, cooperative behaviour
is witnessed in front of the exit, with, e.g., sometimes one participant
waving to a neighbour to go ahead. As the fraction $c_{s}$ of selfish
agents increases, more and more contacts and soft collisions occur,
either on purpose or because of the pressure exerted by the neighbours.
At the same time, the density of participants in front of the exit
rises markedly. At high $c_{s}$, in regions of low density, vying
agents walk visibly faster on average than their polite counterparts
and some align in files of 2-3 selfish agents.

Since selfish agents reach the door faster, they pass more times through
the door in each run. Therefore, the prescribed concentration $c_{s}$
of selfish agents should be replaced by an effective concentration
$c_{s}^{\star}$, corresponding to the number of passages. Table~\ref{tab:summar}
gives the correspondence between the nominal and effective concentrations.

\begin{table}
\begin{centering}
\begin{tabular}{|c | c|c|c|}
\hline 
$c_{s}$ & Effective fraction $c_{s}^{\star}$ & Flow rate $J$ ($\mathrm{s^{-1}}$)
& Density $\rho$ ($\approx\mathrm{m^{-2}}$)\tabularnewline
\hline 
\hline 
0\% & 0\% & $1.26\pm0.07$ & 3.70\tabularnewline
\hline 
10\% & 18\% & $1.39\pm0.09$ & 4.49\tabularnewline
\hline 
60\% & 71\% & $2.20\pm0.09$ & 7.63\tabularnewline
\hline 
90\% & 90\% & $2.36\pm0.15$ & 8.26\tabularnewline
\hline 
100\% & 100\% & $2.41\pm0.17$ & 8.98\tabularnewline
\hline
0\%$^\dagger$ & 0\%$^\dagger$ & $1.01\pm0.05$ & 2.69\tabularnewline
\hline 
30\%$^\dagger$ & 46\%$^\dagger$ & $1.41\pm0.08$ & 4.94\tabularnewline
\hline
60\%$^\dagger$ & 68\%$^\dagger$ & $1.71\pm0.12$ & 6.04\tabularnewline
\hline 
\end{tabular}
\par\end{centering}

\caption{\label{tab:summar}List of the performed evacuation experiments, with
the prescribed fraction of selfish agents $c_{s}$, the effective
fraction $c_{s}^{\star}$, the average flow rate $J$ and the density
$\rho$ in front of the exit. Note that, for $J$, a confidence interval
of $\frac{2\sigma}{\sqrt{n}}$ on either side is given, where $\sigma$
is the standard deviation of $\bar{j}(t)$ averaged over $\delta t=7\,\mathrm{s}$
and $n$ the number of
(independent) sampling points. \newline
$^\dagger$ In these runs, participants were asked to ``head for the door'',
and not to ``head for the door \emph{purposefully}''.}
\end{table}

\subsection{``Instantaneous'' flow rates and the question of stationarity}

\begin{figure}
\begin{centering}
\includegraphics[width=5.4cm]{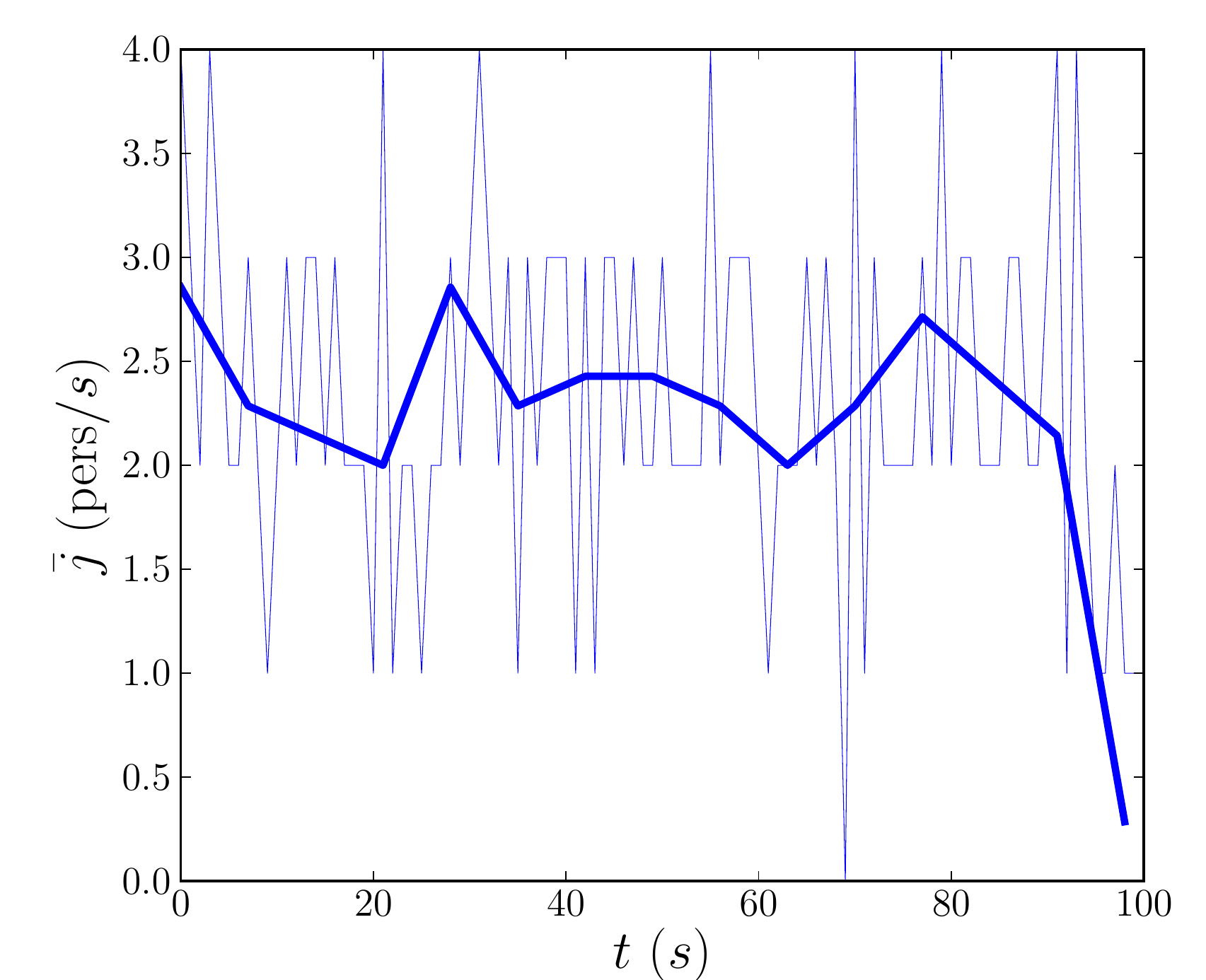}~
\includegraphics[width=5.4cm]{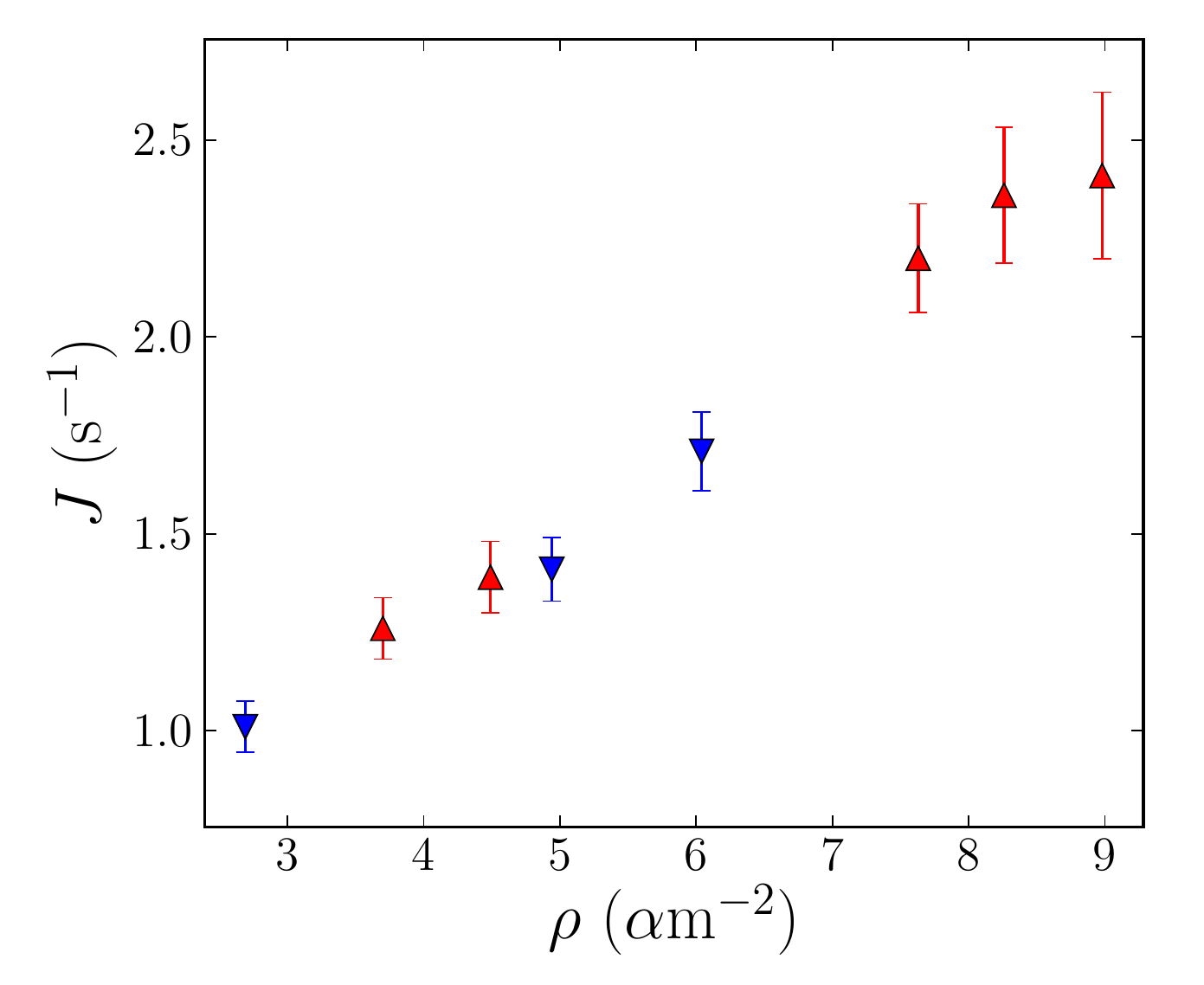}~\includegraphics[width=5.4cm]
{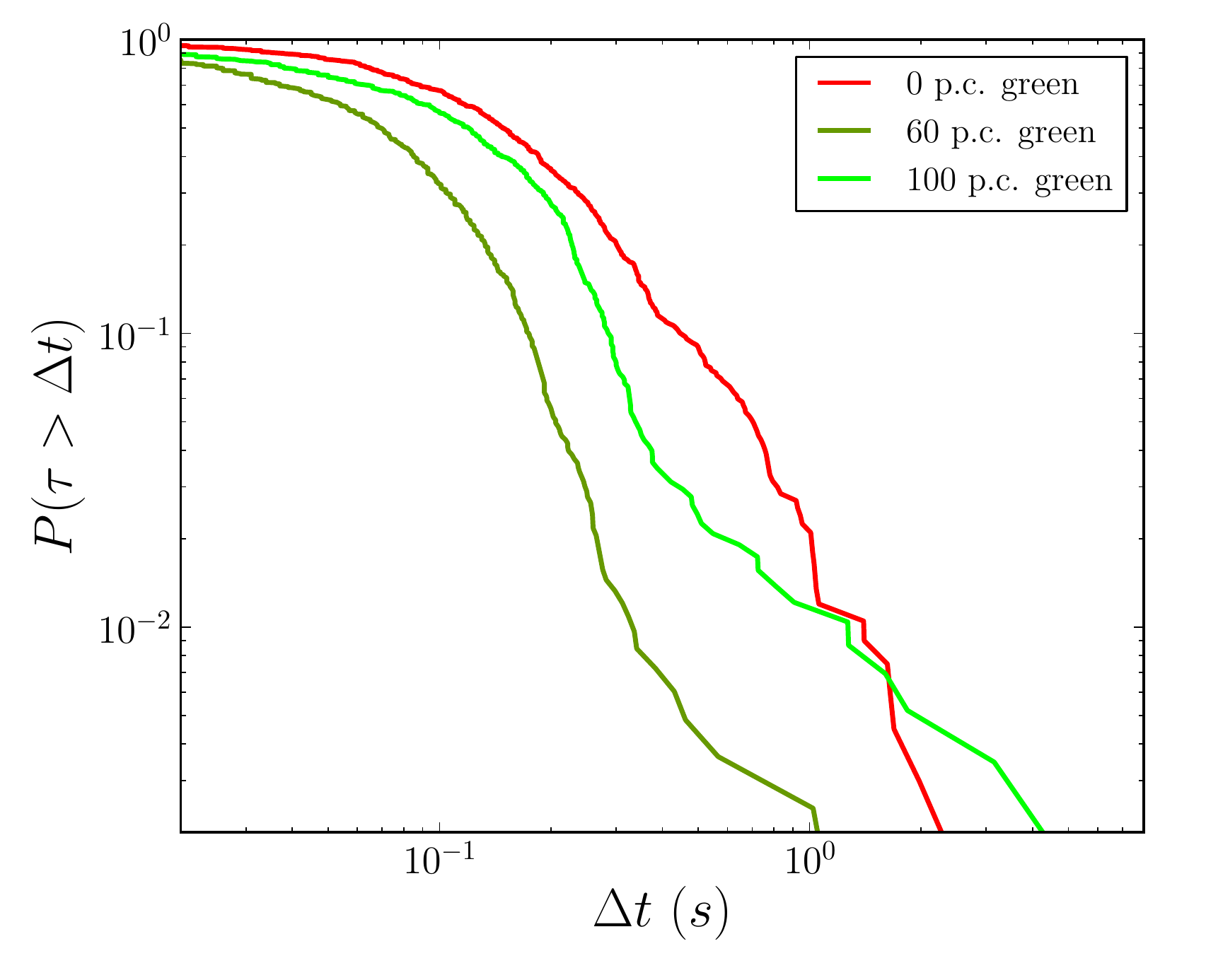}
\par\end{centering}

\caption{\label{fig:J(t)}(\emph{Left}) Time-dependent
flow rate $\bar{j}(t)$ in an experiment involving $c_{s}=90\%$ of
selfish \textbf{pedestrians}. The thin lines are the data averaged
over $\delta t=1\,\mathrm{s}$, whereas the thick lines
are for $\delta t=7\,\mathrm{s}$.
\newline (\emph{Centre}) Dependence of the
average flow rate~$J$ on the \textbf{pedestrian} density $\rho$
in front of the exit. The blue downwards triangles correspond to a set of
evacuation drills in which the participants were just asked to ``head for the
door'' (see footnote in Table~1).
\newline
(\emph{Right}) Complementary cumulated distribution
$P(\tau>\Delta t$) for the \textbf{granular flow }with $w=4\,\mathrm{cm}$;
the data have been cumulated over two to three experiments for each
fraction of non-magnetic (\emph{green}) discs, as indicated in the
legend.}

\end{figure}

To get rid of the dependence on the initial position and of the finite
crowd size effects at the end of the evacuation, it is preferable
to collect data in the stationary state. To ascertain stationarity,
we compute the time-dependent flow rate $\bar{j}(t)$, averaged over
a sliding time window $\delta t$, an example of which is plotted
in Fig.~\ref{fig:J(t)}(left). Clearly, for relatively
small $\delta t$, of order 1s, the flow rate fluctuates considerably.
Averaging over larger time windows, \emph{e.g.}, $\delta t=7\,\mathrm{s}$,
tends to wash out these fluctuations, even though the curve remains
noisy. Nevertheless, beyond this noise, the data clearly point to
the almost immediate attainment of a steady state (for $\bar{j}(t)$),
that persists until (nearly) the end of the experiment. This contrasts
with the bottleneck flow experiments conducted by Seyfried et al.
\cite{Seyfried2009new}, which involved up to 60 participants and
did not quite reach a steady state. However, in their case, the crowd
was initially positioned 3 meters away from the bottleneck and no
recirculation contrivance was used.

Taking a closer look at our time series $\bar{j}(t)$, we realise
that, in the experiment with 10\% of selfish agents in particular,
the flow rate decreases moderately 30 to 40 seconds before the end
of the evacuation (which lasted $190\,\mathrm{s}$ overall). In the
videos, we notice that most selfish agents have already escaped by
this time, hence an excess of polite pedestrians, the last of whom
display no sign of hurry whatsoever.

\subsection{Average flow rate}

Turning to the \emph{average} flow rate $J$, we notice that it increases
monotonically with $c_{s}$ (see Table~\ref{tab:summar}): the more
selfish agents there are, the faster the evacuation. As a corollary
to this statement, no ``faster-is-slower'' effect is observed, unlike
in \cite{Garcimartin2014experimental,Pastor2015experimental}, where
long jams at the door were found to delay significantly the evacuation
for the most competitive crowd. This discrepancy is not necessarily
a contradiction. Instead, we suspect the precise instructions given to the
participants to have mattered: for safety reasons, we insisted repeatedly that
the participants should
not to push each other (although soft contacts were allowed). It follows
that we also observed some jams at the exit for selfish crowds, but
they never lasted long, because there was not enough mechanical pressure
to stabilise them. We shall come back to this issue in the light of
the results for the granular system.

Let us now put in perspective the values we obtained for $J$, which
range from $1.26$ to $2.37\,\mathrm{s^{-1}}$. Using a door of almost
identical width ($69\,\mathrm{cm}$ instead of $72\,\mathrm{cm}$), Pastor
et al. measured global flow rates%
\footnote{Note that $J=\frac{1}{\left\langle \Delta t\right\rangle }$, where
$\left\langle \Delta t\right\rangle $ is the average time lapse between
sucessive escapes in an experiment.%
} $J=2.63\,\mathrm{s}^{-1}$ , $2.56\,\mathrm{s}^{-1}$ and
$2.43\,\mathrm{s}^{-1}$
for homogeneous crowds of moderate, medium, and high competitiveness,
respectively \cite{Pastor2015experimental}. The fact that these values
lie at the upper range of our range of flow rates (corresponding to
$c_{s}=100\%$) supports the idea that the three situations considered
in \cite{Pastor2015experimental} were more competitive than ours,
hence at higher pressure. On the other hand, Seyfried et al. measured
specific capacities $J_{s}$ (defined as the average flow rate per
meter of door width) of about $1.9\,\mathrm{s}^{-1}\cdot\mathrm{m^{-1}}$
in their setup, with a value relatively independent of the bottleneck
width $w$ in the range $[0.9\mathrm{m},1.2\mathrm{m}]$ (in fact,
$J_{s}$ was found to decrease slightly for smaller $w$) \cite{Seyfried2009new}.
For $w=0.72\,\mathrm{m}$,
$J_{s}\lessapprox1.9\,\mathrm{s}^{-1}\cdot\mathrm{m^{-1}}$
yields a flow rate slightly smaller than $1.37\,\mathrm{s}^{-1}$,
which corresponds to the case of our polite crowds ($c_{s}^{\star}=0\%$
or 18\%). Noticing that Seyfried et al.'s experiments were performed
under normal conditions, with no contact between the participants,
we conclude that our results are perfectly consistent with theirs.

\subsection{Density in front of the door}

On the basis of a comprehensive comparison with the literature, Seyfried
et al. claimed that the main parameter controlling the flow rate $J$
was the density $\rho$ in front of the door, while the other parameters
such as the position of the bottleneck or its length were of minor
importance. In spite of their considerable divergences, most models
in the literature, whether empirical or not, predict a quasi-linear
increase of $J$ at low densities, followed by a peak and a decrease
at high densities, due to hampered motion.

\emph{Does this apply to our experiments?} We compute the average
density $\rho$ in a $\approx0.5\mathrm{m}^{2}$ zone in front of
the exit and find values ranging from 4 to 10 $\alpha\mathrm{m}^{-2}$,
approximately, as listed in Table~\ref{tab:summar}. Here, $\alpha$
is a numerical factor close to unity that is mainly due to the optical
distortion of the images; we should also mention that our measurement
method may slightly overestimate the higher range of densities. Going
beyond these caveats, we observe a monotonic increase of $J$ with
the density $\rho$. This holds even for the highest density,
$\rho=9.5\,\alpha\mathrm{m^{-2}}$
at $c_{s}=100\%$, where people are closely packed in front of the
door, whereas the models reviewed in \cite{Seyfried2009new} predict
a decreasing $J\left(\rho\right)$ curve in this range of densities.

Whether the prescribed behaviours affect the flow rate exclusively
indirectly, via the density at the door, or also directly, can only
be settled by a more local analysis of our experiments \cite{nicolas2016next}.
Nevertheless, we would like to remark that at very high densities
the body resistance limits the crowd's compressibility. Accordingly,
an increase of mechanical pressure, \emph{e.g., }due to people pushing
at the back, will only moderately raise the local density, but it
is expected to generate significantly longer clogging events.

\subsection{Is a small fraction of selfish people beneficial for the
evacuation?}

At the 2010 Love Parade in Duisburg, ``pushers'' were recruited
to push the crowd entering the festival area in order to enhance the
inflow \cite{helbing2012crowd}. As a side question, we wonder whether
selfish agents, when they are scarce, act in this way and accelerate
the flow of the lagging polite crowd. Comparing the flows with $c_{s}^{\star}=$
0\% and 18\%, we find that the vying behaviour of some is inefficient
for the rest of the crowd: although their presence increases the global
flow rate $J$, the specific flow rate of polite agents is reduced
with respect to the situation without selfish agents. But more extensive
data at lower values of $c_{s}^{\star}$ would be required to reach
a robust conclusion.

\subsection{Distribution of time lapses between escapes}

\begin{figure}
\begin{centering}
\includegraphics[width=6cm]{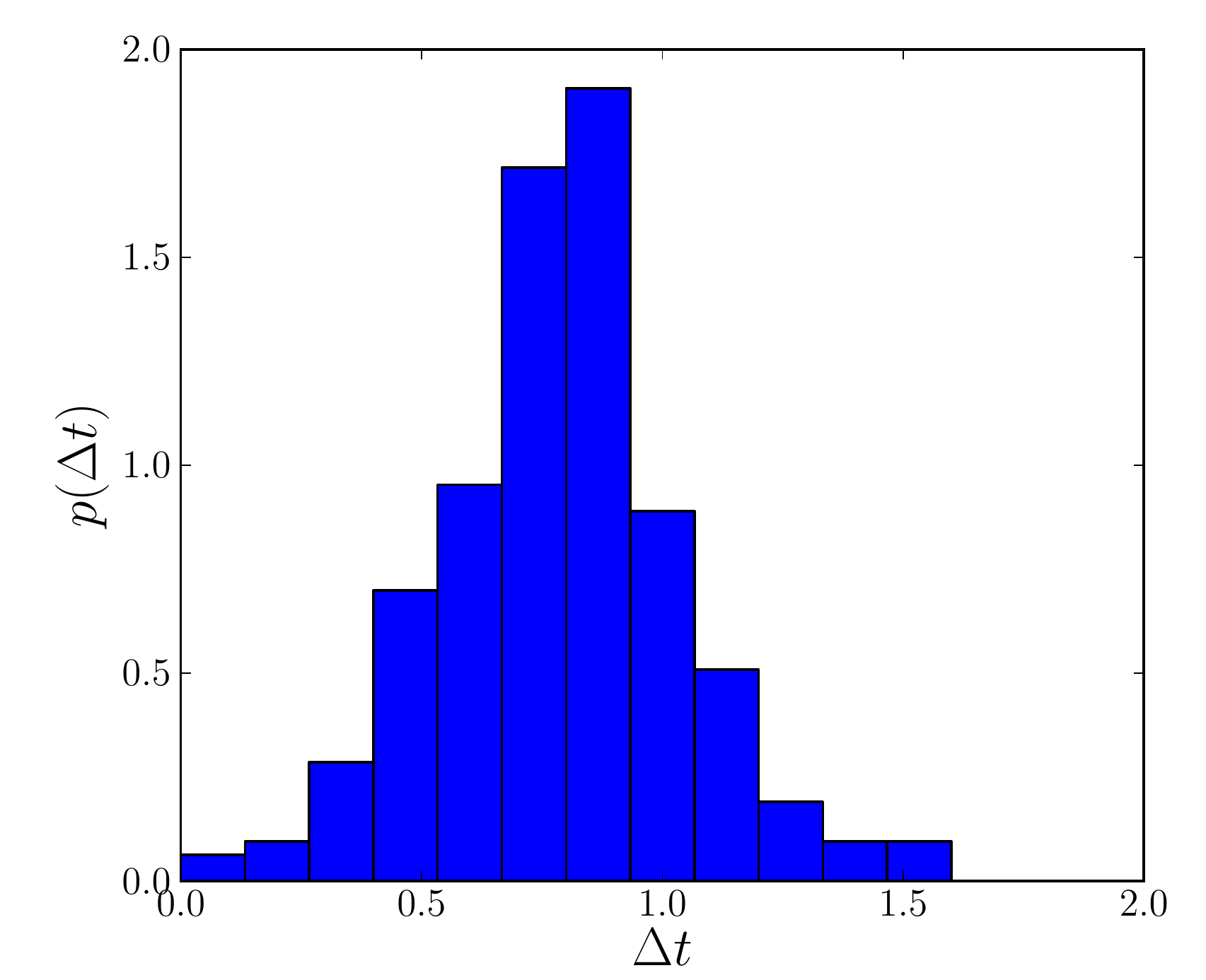}
\includegraphics[width=6cm]{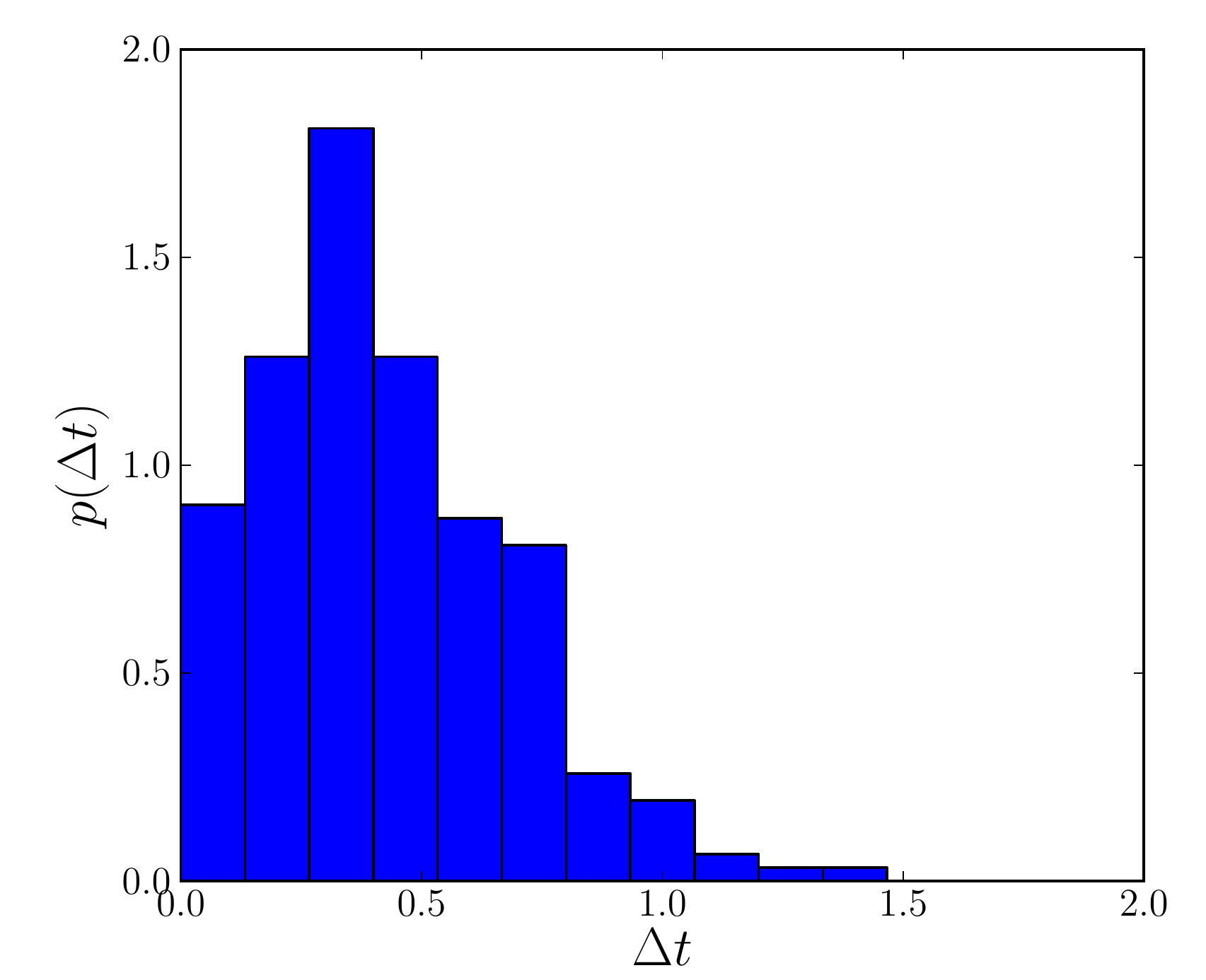}
\par\end{centering}

\caption{\label{fig:Histograms-of-time}Histograms of time lapses $\Delta t$
between successive escapes in the experiment with $c_{s}^{\star}=0\%$
 of selfish agents (\emph{left}) and $c_{s}^{\star}=90\%$ (\emph{right}).}

\end{figure}

In Fig.~\ref{fig:Histograms-of-time}, we plot the distributions
of time lapses $\Delta t$ between successive escapes through the
door for $c_{s}^{\star}=0\%$ and 90\%. For the polite crowd, the
distribution is peaked at a value of $\Delta t$ slightly below $1\,\mathrm{s}$.
This peak is shifted to lower value for $c_{s}^{\star}=90\%$ and,
above all, the distribution broadens and the probability of very small
values $\Delta t\approx0$ is strongly enhanced. This feature turns
out to be characteristic of flows with a significant fraction $c_{s}^{\star}$
of selfish agents and reflects bursts of quasi-simultaneous escapes
through the door.

\section{Experimental results for the granular flow}

\subsection{General observations; (non)-stationarity}

Having analysed the global statistics of pedestrian evacuations through
a narrow door, we turn to the granular flow. In systems of only magnetic
discs ($c_{s}=0\%$, with $c_{s}$ now referring to the fraction of
non-magnetic grains), the density increases gradually, but quite visibly,
with the depth: at the bottom, the discs are much closer to each other
than at the top, but still non-touching. This separation gets larger
as the experiment proceeds and the weight of the granular layer decreases.
In other words, there is no Janssen effect. It also follows that the
flow is non-stationary at $c_{s}=0\%$: the flow rate $\bar{j}(t)$
slows down with time (\emph{data not shown}). Besides, no persistent
clog is observed, even for the narrowest door, $w=3\,\mathrm{cm}$.
We ascribe this difference with respect to Lumay et al.'s recent
experiment with magnetic discs \cite{lumay2015flow} to the absence
of vibrations in their setup.

On the other hand, in systems with a significant fraction of non-magnetic
discs ($c_{s}=60\%$, 100\%), persistent clogs occur for $w=3\,\mathrm{cm}$,
which have to be broken manually. For larger doors, the vibrations
suffice to unclog the system. Besides, there is not so prominent a
decrease of the flow rate $\bar{j}(t)$ with time and a quasi-stationary
regime seems to be reached for $c_{s}=100\%$ (as expected
in a non-vibrated system), even though fluctuations blur the picture.

\subsection{Average flow rates}

For the narrow door, $w=3\,\mathrm{cm}$, magnetic grains flow faster
overall than neutral ones, which undergo severe clogs.
It is tempting to draw a parallel with the pedestrian evacuation experiments
of \cite{Pastor2015experimental}, in which the collision-prone competitive
crowd evacuated more slowly than the less competitive one. We should
however remark that, here, the isolated neutral discs do not slide
faster than their magnetic counterparts.

The situation differs for $w=4\,\mathrm{cm}$. In this case, the global
flow rate $J$ is larger for the system which exhibits contacts and
collision, \emph{i.e.}, here, the non-magnetic one ($J=6.2\,\mathrm{s^{-1}}$
at $c_{s}=100\%$ \emph{vs.} $J=4.7\,\mathrm{s^{-1}}$ at $c_{s}=0\%$).
This echoes our observations in the pedestrian experiments. Nevertheless,
contrary to these, the flow rate $J$ for an intermediate
fraction of non-magnetic discs, $c_{s}\simeq60\%$, is by far larger
($J\approx11\,\mathrm{s^{-1}}$) than that at $c_{s}=100\%$; the non-stationarity
of the flow at $c_{s}=0\%$ makes another difference.

\subsection{Time lapses}

Figure~\ref{fig:J(t)}(right) presents the complementary
cumulated distribution $P(\tau>\Delta t)$ of time lapses between
successive egresses.

As in the pedestrian experiment, the more negative slope of $P(\tau>\Delta t)$
at small $\Delta t$ indicates that there are more quasi-simultaneous
egresses at large $c_{s}$ than at $c_{s}=0\%$, where grains are
separated. On the other hand, the softening of the slope at large
$\Delta t$ reveals the presence of long clogs at large $c_{s}$,
which are indeed observed in the videos. This is in line with the
stronger flow intermittency in the presence of many vying
pedestrians.

\section{Discussion and outlook}

We have studied the influence of polite \emph{vs.} selfish behaviours
on the evacuation of pedestrians through a narrow door. We have observed
a monotonic increase of the global flow rate with the fraction $c_{s}$
of selfish agents, \emph{i.e.}, a ``faster-is-faster'' effect, and
with the density $\rho$ in front of the door, up to
$\rho\approx9-10\,\mathrm{m^{-2}}$.
In parallel, evacuations at high $c_{s}$ display stronger intermittency
and are characterised by the presence of quasi-simultaneous escapes.
Under the hypothesis that the absence of mutual contacts characterise
the motion of polite pedestrians, we have proposed to model them as
repulsive magnetic discs in a granular hopper flow. This analogy helps
us explain some features observed in the pedestrian evacuation in
mechanical terms, but also suffers from some deficiency, such as the
absence of stationarity in the contact-free flow.

\emph{Acknowledgements.} We are grateful to the Grupo de Higiene y Seguridad of
the CAB for their help in devising the pedestrian
experiment and to all the voluntary participants.

\end{document}